\documentclass[12pt, prd, showpacs]{revtex4}
\usepackage{amssymb}
\usepackage{amsmath}
\usepackage{graphicx}

\begin{document}

\title{High energy particle collisions under Cauchy horizon}
\author{A. V. Toporensky}
\affiliation{Sternberg Astronomical Institute, Lomonosov Moscow State University, 119991
Moscow, Russia}
\email{atopor@rambler.ru}
\author{O. B. Zaslavskii}
\affiliation{Department of Physics and Technology, Kharkov V.N. Karazin National
University, 4 Svoboda Square, Kharkov 61022, Ukraine}
\email{zaslav@ukr.net }

\begin{abstract}
We consider particle collision inside the inner horizon of the
Reissner-Nordstrom metric in the so-called R region. We show that there
exist scenario in which the enrgy in the center of mass frame grows
unbounded. In contrast to the standard scenarios of high energy collisions
in black hole background in the R region, fine tuning of particle parameters
is not require. The effect found in this work can be considered as a massive
particle counterpart of wave processes that contribute to instability of the
inner black hole horizon.
\end{abstract}

\maketitle

\section{Introduction}

The original Ba\~{n}ados-Silk-West (BSW) effect consists in unbounded growth
of the energy in the center of mass frame $E_{c.m.}$ when two particles
collide near the event horizon of a black hole \cite{ban}. In doing so, one
of particles should have fine-tuned parameters. Such an effect was found for
particle collisions outside the extremal event horizon in the R region
(following classification of \cite{novrt}). Later on, a similar effect was
found for the static charged black holes \cite{jl}. Further, collisions
inside the event horizon between the Cauchy and event \ horizon of the
nonextremal Reissner-Nordstr\"{o}m (RN)\ black hole in the T region were
discussed and the possibility of unbounded growth of $E_{c.m.}$ was
established (see \cite{inner} and references therein to preceding discussion
which was controversial). The aim of this Letter is to show that there
exists also another kind of high energy particle collisions which are
possible also under the Inner (Cauchy) horizon in the R region of the RN
metric. To the best of our knowledge, this kind of the effect was not
considered in literature before.

\section{Basic equations}

Let us consider the RN metric%
\begin{equation}
ds^{2}=-fdt^{2}+\frac{dr^{2}}{f}+r^{2}(d\theta ^{2}+\sin ^{2}\theta d\phi
^{2})\text{,}  \label{met}
\end{equation}

where%
\begin{equation}
f=1-\frac{2M}{r}+\frac{Q^{2}}{r^{2}}\text{,}
\end{equation}%
$M$ is the black hole mass, $Q$ being its electric charge. We use the system
of units in which fundamental constants $G=c=1$. If $M>Q$, there exists the
event horizon at $r_{+}=M+\sqrt{M^{2}-Q^{2}}$ and the inner (Cauchy) horizon
at $r_{-}=M-\sqrt{M^{2}-Q^{2}}$.

For simplicity, we restrict ourselves by pure radial motion and assume that
particles are neutral, so they follow geodesic paths. The equations of
motion of a particle having mass $m$ read%
\begin{equation}
m\dot{t}=\frac{E}{f}\text{,}  \label{te}
\end{equation}%
\begin{equation}
m\dot{r}=\sigma P\text{, }P=\sqrt{E^{2}-m^{2}f}\text{,}  \label{r}
\end{equation}%
where $\sigma =\pm 1$, dot denotes derivative with respect to the proper
time. $E$ is the energy that conserves due to the existence of the time-like
Killing vector. Correspondingly,%
\begin{equation}
t=\int_{r_{0}}^{r}\frac{dr^{\prime }E}{f(r^{\prime })P(r^{\prime })}+t_{0}%
\text{,}  \label{t+}
\end{equation}%
if $\dot{r}>0$ and%
\begin{equation}
t=-\int_{r_{0}}^{r}\frac{dr^{\prime }E}{f(r^{\prime })P(r^{\prime })}+t_{0}%
\text{,}  \label{t-}
\end{equation}%
if $\dot{r}<0$. It is assumed that $r=r_{0}$ at $t=t_{0}$.

In eqs. (\ref{met}) - (\ref{r}) the coordinate $r$ is space-like and $t$ is
time-like. This is true in the $R$ region which exists for $r>r_{+}$ and $%
0<r<r_{-}$. We will concentrate just on the second possibility.

\section{Energy in the center of mass and properties of motion}

If two particle collide in some point, the energy $E_{c.m.}$ in the center
of mass, by definition, is equal to 
\begin{equation}
E_{c.m.}^{2}=-P_{\mu }P^{\mu }\text{,}
\end{equation}%
where $P^{\mu }=m_{1}u_{1}^{\mu }+m_{2}u_{2}^{\mu }$ is the total
energy-momentum. Then,%
\begin{equation}
E_{c.m.}^{2}=m_{1}^{2}+m_{2}^{2}+2m_{1}m_{2}\gamma \text{,}
\end{equation}%
where $\gamma =-u_{1\mu }u_{2}^{\mu }$ is the Lorentz factor of relative
motion. It follows from (\ref{te}), (\ref{r}) that%
\begin{equation}
\gamma =\frac{E_{1}E_{2}-\sigma _{1}\sigma _{2}P_{1}P_{2}}{f}\text{.}
\end{equation}

Now, we consider this quantity for the metric (\ref{met}). When $%
r\rightarrow 0$, $f\rightarrow \infty $. It means that in the inner R region
the turning point is inevitable. If $\sigma =-1$, 
\begin{equation}
\gamma =\frac{E_{1}E_{2}+P_{1}P_{2}}{f}\text{.}  \label{ga}
\end{equation}%
The idea, \ how one can arrange high energy particle collision, consists in
the following. Particle 1 enters the R region, turns back and meets particle
2 that moves with the negative radial velocity. Then, $\sigma _{1}=+1$ and $%
\sigma _{2}=-1$. Therefore, eq. (\ref{ga}) applies and we obtain that $%
\gamma \rightarrow \infty $ if collision happens as close to the horizon as
we like.

\section{Kinematics}

However, this is not the end of story. The problem to arrange a suitable
scenario includes not only an unbounded growth of $\gamma $ but also
kinematic layout that makes collision possible. Otherwise, there is no event
itself, so eq. (\ref{ga}) becomes meaningless. Here, there exists
difficulties similar to those for collisions between the horizons \cite%
{inner}. Namely, for typical paths, particles 1 and 2 cross different
branches of the inner horizon, and there is no event of collision at all.
(See Fig. 1).

\begin{figure}[tbp]
\centering
\includegraphics[width=0.7\linewidth]{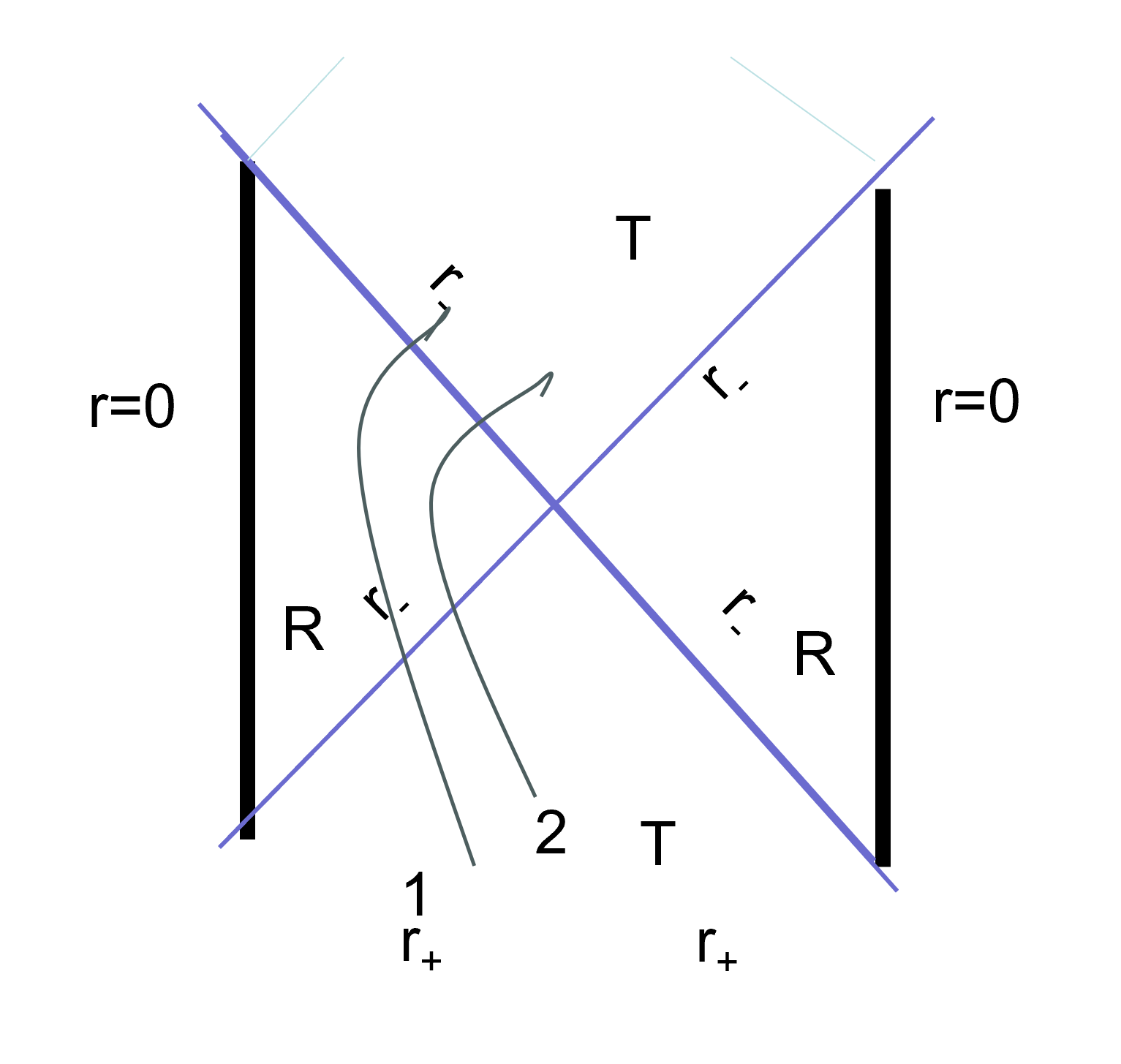}  
\caption{No particle collision}
\label{fig1.png}
\end{figure}

However, we can achieve collision by a proper choice of the constants of
integration. In the point of collision $r_{c}$ of particles 1 and 2 we must
have $r_{1}=r_{2}=r_{c}$, $t_{1}=t_{2}$. Thus we have from (\ref{t+}) and (%
\ref{t-}) that%
\begin{equation}
\int_{r_{0}^{(1)}}^{r}\frac{dr^{\prime }E_{1}}{f(r^{\prime })P_{1}(r^{\prime
})}+t_{0}^{(1)}=-\int_{r_{0}^{(2)}}^{r}\frac{dr^{\prime }E_{2}}{f(r^{\prime
})P_{2}(r^{\prime })}+t_{0}^{(2)}\text{,}
\end{equation}%
so%
\begin{equation}
t_{0}^{(2)}-t_{0}^{(1)}=\int_{r_{0}^{(1)}}^{r_{c}}\frac{dr^{\prime }E_{1}}{%
f(r^{\prime })P_{1}(r^{\prime })}+\int_{r_{0}^{(2)}}^{r_{c}}\frac{dr^{\prime
}E_{2}}{f(r^{\prime })P_{2}(r^{\prime })}\text{.}  \label{t21}
\end{equation}

We try to arrange such a scenario in which $r_{c}$ is close to $r_{-}$. It
means that one of particles has been continuing to move from the horizon but
does not deviate from it too far, so it remains in the near-horizon region.
Then, it follows from (\ref{t21}) that for $r_{c}\rightarrow r_{-}$, where $%
f\rightarrow 0$,%
\begin{equation}
t_{0}^{(2)}-t_{0}^{(1)}\approx 2\int_{r_{c}}^{r-}\frac{dr^{\prime }}{%
f(r^{\prime })}
\end{equation}

Using expansion 
\begin{equation}
f\approx 2\kappa _{-}(r-r_{-})\text{,}
\end{equation}%
where $\kappa _{-}=\frac{f^{\prime }(r_{-})}{2}$ is the surface gravity of
the inner horizon, we obtain that%
\begin{equation}
t_{0}^{(2)}-t_{0}^{(1)}\approx \frac{1}{\kappa _{-}}\ln \left\vert
r_{c}-r_{-}\right\vert
\end{equation}

This quantity diverges in the limit $r_{c}\rightarrow r_{-}$. For the left
hand side of this equation to be unbounded from above, we need either $%
t_{0}^{(1)}\rightarrow -\infty $ (particle 1 starts its motion in an almost
infinite past, see Fig.2)

\begin{figure}[tbp]
\centering
\includegraphics[width=0.7\linewidth]{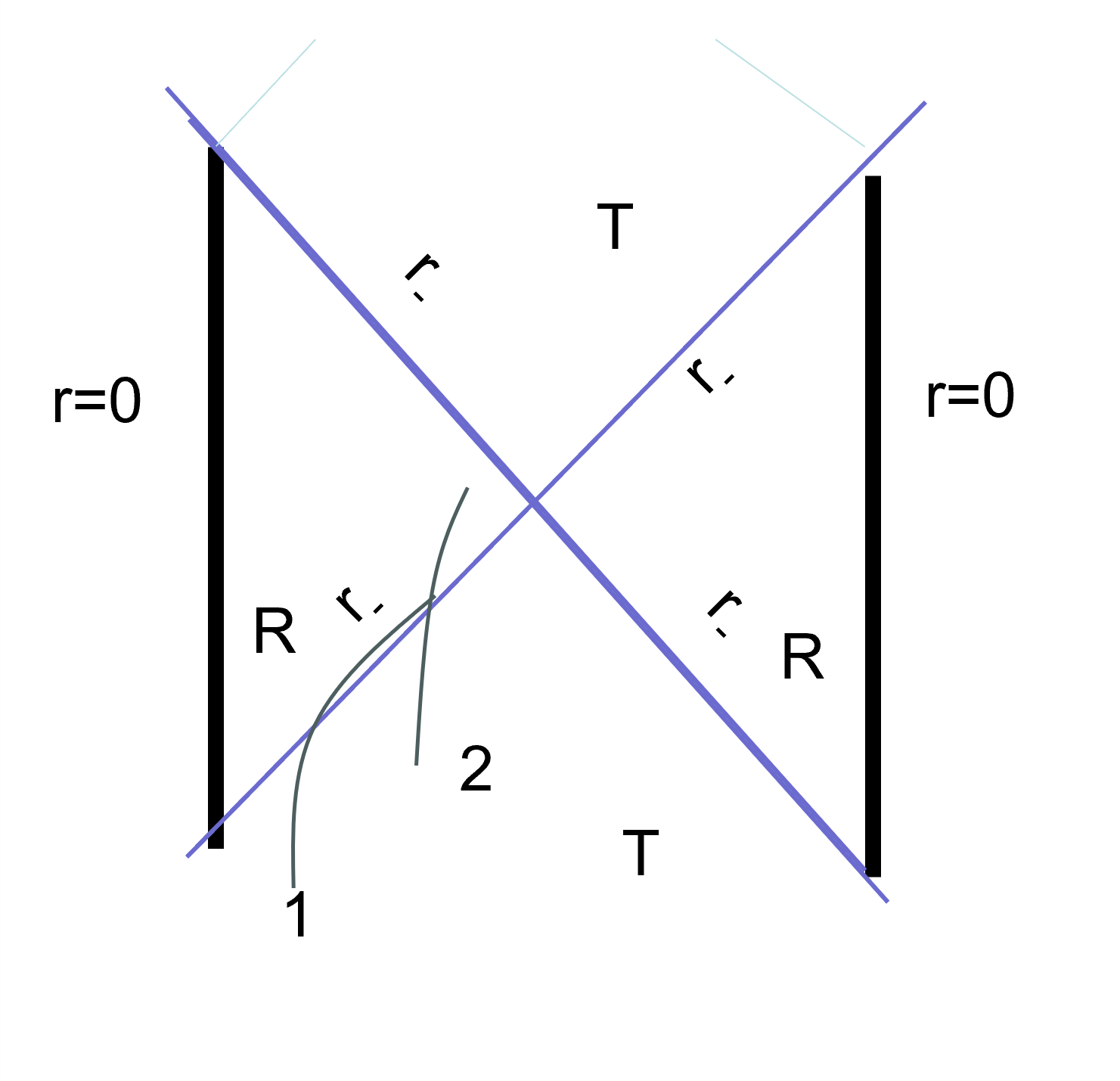}  
\caption{Particle 1 starts its motion in far past}
\label{figure2_newnew.png}
\end{figure}

or $t_{0}^{(2)}\rightarrow +\infty $ (particle 2 approaches the future
horizon in an almost infinite future, see Fig.3).

\begin{figure}[tbp]
\centering
\includegraphics[width=0.7\linewidth]{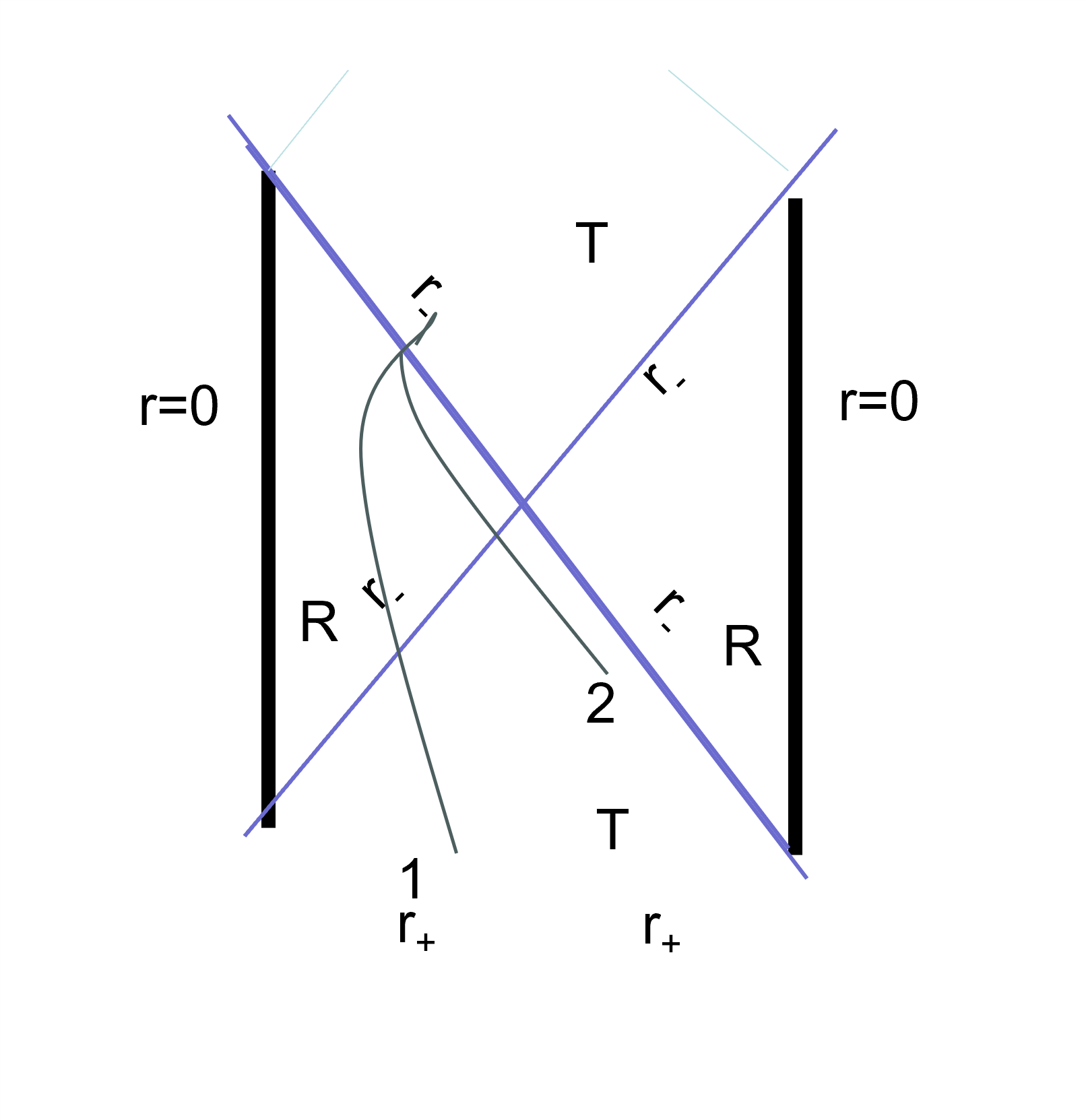}  
\caption{Particle 2 approaches horizon in far future}
\label{figure_3newnew.png}
\end{figure}

\section{Conclusion}

Thus we found one more type of scenarios of particle collisions that lead to
an infinite growth of $E_{c.m.}$ It happens (i) inside the inner horizon in
the R region, (ii) fine-tuning of one of particles, typical of the standard
BSW effect, is not required here. Both particles are usual in this sense. We
considered neutral particles but the effect under discussion should exist
also for particle with electric charge, with arbitrary relation between the
energy and charge.

Meanwhile, the results obtained are not only relevant on their own. It is
general belief that the inner black hole horizon is unstable \cite{isr}. The
arguments rely on the behavior of wave packets in the region between the
inner and outer horizon. In the present work, we considered the region
beyond the inner horizon and discussed collisions of massive particles. In
this sense, our work is complementary to previous ones both in what concerns
the region under investigation and in that we analyzed motion of massive
particles instead of radiation. Since the collision energy appears to be not
bounded from above, we deem that this effect can contribute to the
instability of the inner black hole horizon, though this particular problem
needs further studies.

There is one more important point. In any event, the energy cannot be
literally infinite, although it can be as large as one likes. This is the
contents of what is called kinematic censorship \cite{cens}. In our case it
is fulfilled since its violation would require infinite $t_{0}$ and motion
of a massive particle along the light-like surface of the horizon that is
impossible.

Although we restricted ourselves by static electrically charged black holes,
it is clear that the similar phenomenon should exist for rotating black
holes.

Thus the BSW effect or its analogue near the black hole horizon is
established for all possible configurations: near the event horizon outside
it \cite{ban}, \cite{jl}, between horizons inside the event horizon \cite%
{inner} and inside the inner horizon. In this sense, we completed
classification of scenarios relevant for the BSW effect.

\end{document}